\documentclass[prd,superscriptaddress,a4paper,showpacs,showkeys,nofootinbib]{revtex4}
\usepackage{graphicx}%  Default Latex 2eps package for embedding figures
\usepackage{subfigure}
\usepackage{rotating}
\usepackage{epsfig}
\usepackage[lofdepth,lotdepth,caption=false]{subfig}
\usepackage{bm}
\usepackage{amsmath}
\usepackage{dcolumn}% Align table columns on decimal point
\usepackage{bm}% bold math

\usepackage{url}
\usepackage{amsmath,amssymb,graphicx}
\usepackage{amsmath}
\usepackage{amssymb}
\usepackage{mathrsfs}
\usepackage{accents}
\usepackage{epsfig}
\usepackage{enumerate}

\usepackage{mathtools}
\usepackage{mathrsfs}

\usepackage{color}
\usepackage{bm}
\usepackage{amsfonts}
\usepackage{amssymb,amscd}

\def\lsim{\raise0.3ex\hbox{$<$\kern-0.75em\raise-1.1ex\hbox{$\sim$}}}

\def\gsim{\raise0.3ex\hbox{$>$\kern-0.75em\raise-1.1ex\hbox{$\sim$}}}

\newcommand{\be}{\begin{equation}}

\newcommand{\ee}{\end{equation}}

\def\beq{\begin{equation}}

\def\eeq{\end{equation}}

\def\beqa{\begin{eqnarray}}

\def\eeqa{\end{eqnarray}}

\newcommand{\ba}{\begin{eqnarray}}

\newcommand{\ea}{\end{eqnarray}}

\def\gappeq{\mathrel{\rlap {\raise.5ex\hbox{$>$}}

{\lower.5ex\hbox{$\sim$}}}}

\def\lappeq{\mathrel{\rlap{\raise.5ex\hbox{$<$}}

{\lower.5ex\hbox{$\sim$}}}}

\def\Toprel#1\over#2{\mathrel{\mathop{#2}\limits^{#1}}}

\begin{document}

%\title{Investigating the impact of the polarization of the Earth's outer core on the neutrino absorption}
\title{Investigating the impact of spin effects at the high-energy neutrino-nucleon interactions while it crosses the Earth's core}
\author{R. Francener}
\affiliation{Instituto de Física Gleb Wataghin - UNICAMP, 13083-859, Campinas, SP, Brazil. }
\author{D.R. Gratieri}
\affiliation{Escola de Engenharia Industrial Metalúrgica de Volta Redonda, Universidade Federal Fluminense (UFF), 27255-125, Volta Redonda, RJ, Brazil}
\affiliation{Instituto de Física Gleb Wataghin - UNICAMP, 13083-859, Campinas, SP, Brazil. }
\author{G. Torrieri}
\affiliation{Instituto de Física Gleb Wataghin - UNICAMP, 13083-859, Campinas, SP, Brazil. 
}
\affiliation{Institute of Physics, Jan Kochanowski University, Ul. Uniwersytecka 7, 25-406 Kielce, Poland. 
}

\begin{abstract}
In this work, we investigate the impact of assuming the
polarization of the Earth’s outer core on the propagation of neutrinos that cross the all the Earth. 
We taking into account the spin-dependent structure functions to describe the polarized
neutrino-nucleon cross-section, and also on the neutrino absorption while it crosses all of the Earth.
We found that adding spin information and simultaneously assuming polarization of Earth’s outer
core impacts the probability of neutrino absorption in the energy range of 10 - 100 TeV and for
upward neutrino direction. However, the magnitude of the effect is small and should be comparable
with the magnitude of the errors associated with the IceCube neutrino data.

\end{abstract}



\maketitle

\vspace{1cm}

\section{Introduction}
\begin{figure}[t]
 \includegraphics[scale=0.45]{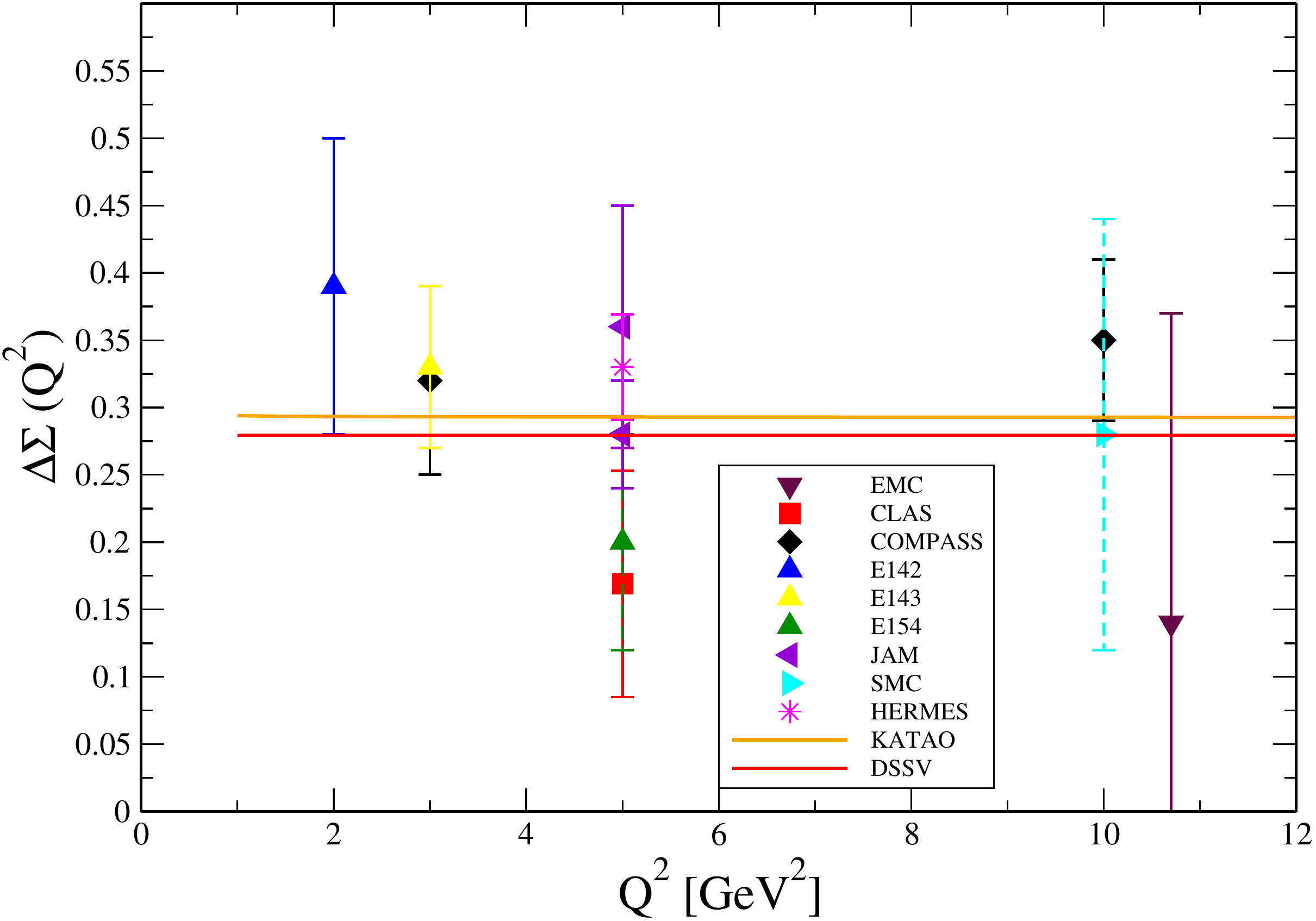} 
\caption{ Compilation of experimental results for the quarks contribution  to proton spin from \cite{EMC_1988, EMC_1989,COMPASS,COMPASS3, compass2, clas, hermes2, SMC, e143, e142, e154, jlab}. For comparison, the prediction from KATAO \cite{KATAO,KATAO2,KATAO3} and DSSV \cite{florian,florian2}  are also shown. 
}
\label{fig:p_Spin_Picture}
\end{figure}

Interactions with polarized nuclei have gained great attention in theoretical and experimental physics community  since the results of the  {\it European Muon Collaboration (EMC)} from the late eighties  \cite{EMC_1988,EMC_1989}.   Such results pointed out that {\it ``the total quark spin constitutes  only a small  fraction $\Delta\Sigma(Q^{2})$ of the proton's spin"}.  This result became known as the {\it ``Proton Spin Crisis" (PSC)}. Usually, the nucleon spin is assumed to be given in terms of the sum of the spin contributions from quarks, gluons and orbital magnetic moment from both quarks and gluons, 
\begin{eqnarray}
    \frac{1}{2} = \frac{1}{2}\Delta\Sigma (Q^{2})+ \Delta G(Q^{2}) + L_{q}(Q^{2})+L_{g}(Q^{2}),
\label{Eq:PProblem}
\end{eqnarray}
where $\Delta G(Q^{2})$  is the gluon  contribution to the nucleon spin,  $L_{q}(Q^{2})$ and $L_{g}(Q^{2})$ are the {\it Orbital Angular Momentum (OAM)} contributions from quarks and gluons, respectively. The quark contribution to nucleon spin, $\Delta \Sigma(Q^{2})$, can also be understood in terms of the sum in all quarks of integral in \textit{x} of helicity distributions $(\Delta q^{i}(x,Q^{2}))$ \cite{EMC_1988,EMC_1989}. In all cases, in the {\it Naive Parton Model}, the quantity $\Delta q^{i}(x,Q^{2})dx $ is the number of polarized (anti)quarks of type $q$ carrying a momentum fraction between $x$ and $x+ dx$. The index $i = u,d,c,s,b,t$ stands for each quark flavor. Within the {\it Quantum Cromodynamics (QCD)} there are both perturbative and non-perturbative corrections, in such way that the structure functions associated with the nucleons are obtained from the partonic density functions through the {\it Factoriazation Theorem} \cite{Collins:1989gx}. For a complete review, see  \cite{Workman:2022ynf}. For the theoretical formalism, we point to reference \cite{pol}, which we follow closely. See also \cite{pol2}. Explicitly, the result from the EMC collaboration  pointed out that $\Delta \Sigma(Q^{2}) = 0.14 \pm 0.23$.  Current  analysis from  COMPASS collaboration \cite{COMPASS} report that  about $31\% \pm 11\%$ of proton spin comes from quarks, for $Q^2 = 3\; \mathrm{GeV}^2$.  Actually, the literature seems to converge for the longitudinally spin polarization scattering, but there still a puzzle for  transversely polarized nuclear targets interacting with transversely polarized  projectile nucleons ($\approx 40\%$) and  charged leptons ($\approx 5-10\%$) \cite{Spin_quark_sea}.  To illustrate the actual scenario, in  Fig. \ref{fig:p_Spin_Picture}  we present the contribution to proton's spin due to quarks measured  by several collaborations \cite{EMC_1988,EMC_1989, COMPASS,COMPASS3, compass2, clas, hermes2, SMC, e143, e142, e154, jlab} as function of the respective momentum scale, $Q^{2}$. For comparison, our results assuming the predictions from  \cite{KATAO,KATAO2,KATAO3,florian,florian2} 
% and  \cite{florian,florian2}%
 are also shown.  At the present level of experimental accuracy, the scalling of polarized PDFs  was still not clearly seen \cite{Workman:2022ynf}. In fact,  the structure functions associated with the  quarks are relatively the most known ones for the both unpolarized and polarized cases. Moreover, a recent work from {\it JAM Collaboration} \cite{Cocuzza:2022ovz}  based on the  STAR  data \cite{Ethier:2017zbq} in the range of $0.01 \le x \le 0.3$ and $Q^{2} = 10 $ GeV $^{2}$, presents for the first time results favoring the nonzero helicity sea assymmetry, with $\Delta {\cal{X}}^{2}/N_{dat} \approx 1\sigma$. Such results are from a global analysis of both unpolarized and polarized PDFs. Concerning the gluon spin, we  know today  \cite{Workman:2022ynf}  that at small $x$ and large $Q^{2}$ the {\it gluon density function, $g(x,Q^{2})$ },  is considerably larger than the density functions associated to the quarks, which implies that at the high energy regime the nucleon can be understood as a collective of gluons. Hence, it is straightforward to expect some degree of contribution from  gluons to the nucleon spin. Indeed, in \cite{Spin_quark_sea}, which is a review of the topic, it is stated that while the contribution from the valence quarks saturates at the high-energy  limit,  the gluon contribution is expect to reach $\approx 50\%$ at the present accelerator energies. This is in agreement with recent results from lattice QCD \cite{lattice} and also and experimental analyzes \cite{exp_gluon}.  Moreover, since $\Delta\Sigma(Q^{2})$  and  $\Delta G(Q^{2})$ are observables,  overall spin conservation  $(S_{p} = 1/2)$ can be used, and implies in large contribution from OAM to the proton spin \cite{Scopetta:1999ue}. Indeed, in  \cite{Ratcliffe} it is shown that orbital angular momentum is generated in the partonic dynamical evolution as it is given by the {\it DGLAP} equations \cite{Gribov:1972ri,Altarelli:1977zs,Dokshitzer:1977sg}.

To better describe the quark and gluon content and aspects of the three-dimensional structure of the nucleon, there are generalized parton distribution (GPDs) and transverse momentum dependent distributions (TMDs) \cite{introduction_gpd,introduction_gpd2}. Such distributions are complementary and aim to describe the transverse plane of nucleon propagation. Another important class of parton distributions are the longitudinal spin-dependent ones, which describe the asymmetry between quarks with opposite spins in the nucleon.
Recently, several experimental collaborations have focused their efforts on measuring the (longitudinal) spin-dependent structure functions in collisions of charged leptons with polarized hydrogen, deuterium and helium-3 nuclei \cite{e143,hermes,compass2,clas,jlab}. Current measurements focus on modest kinematic ranges ($x > 10^{-3}$ and $Q^2 < 10^2\; \mathrm{GeV}$). With existing data on spin asymmetry and spin-dependent structure functions, different authors have built fits that parameterize these data \cite{florian,florian2,KATAO,KATAO2,KATAO3} and allow extrapolations beyond observed kinematic ranges.
One of the main goals of the future Electron-Ion Collider (EIC) \cite{eic} is to improve our understanding of the helicity distributions of quarks and gluons inside nucleons and heavy nuclei. With this new collider it is intended to increase the current observation range of these distributions. While $x$ should be decreased to approximately $5\cdot 10^{-5}$, $Q^2$ should be increased to approximately $10^3\; \mathrm{GeV}^2$ (see Fig. 10 in \cite{eic}). 
Typical IceCube events occur with $x\sim  10^{-2,\; -3}$ and $Q^2 \sim 10^{3,\; 4}$, very close to the future EIC data, making the necessary extrapolation much closer and with a high confidence level.
 %Olhar Ref PDG 2022}
 
% but there is a proposal to build a large Electron-Ion Collider \cite{eic} that should expand the observation range, as well as make measurements with ions that have not yet been tested. 

%\vspace{1,5cm}
%Aqui introduzimos TMDs SI-DIS, etc e como estas se relacionam com o problema em quest\~ao.

%Nucleon parton distributions cannot be obtained by first principles of quantum chromodynamics \cite{halzen}, which makes them totally dependent on parameterizations of experimental measurements. In the parton model, nucleons are described by distributions of their constituents in terms of the fraction of nucleon momentum carried by the quasi-free parton, $x$, and the square of the four-momentum transferred by it, $-Q^2$. 

%\vspace{1,5cm}
%Aqui diremos o que está sendo planejado para o futuro próximo, o experimento EIC e qual o seu  range em x, Q2. Também seria interessante colocar uma conexão com lattice-QCD 

%\vspace{0,5cm}
%Aqui temos que amaciar a transição collider para observatório de neutrinos, dizer porque o IceCube pode ser sensível a PDFs em uma região de x e Q2 ainda não estudada, que vamos testar isso e como vamos fazer. Importante citar trabalhos que estudam interação forte e neutrinos e trabalhos que usam neutrinos para estudar o interiror da Terra para valorizar o nosso trabalho. 

\begin{figure}[t]
	\centering
\includegraphics[width=0.35\textwidth]{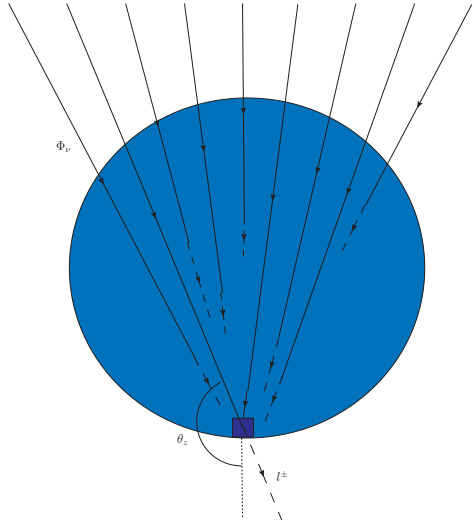} 
	\caption{ Attenuation of neutrino flux by Earth's absorption.}
	\label{fig:diagram}
\end{figure}

In this work we present a study of the impact of polarization of hadronic targets on Deep Inelastic Scattering (DIS) of muonic neutrinos and antineutrinos, and we apply the obtained result to the neutrinos absorption by the Earth. Such a study is strongly motivated by a probable polarization of the Earth's outer core, whith is generated by a turbulent flow of liquid metal \cite{geodynamics}. We will verify if through the interaction of neutrinos with the Earth it is possible to estimate the polarization of the outer core.
This study is also motivated by the recent IceCube measurement of the cross section of muonic neutrinos by the Earth's absorption \cite{measure}. This measurement indicates that the cross section, in the observed energy range ($6.3 - 980\; \mathrm{TeV}$), is about 1.3 times the cross section predicted by the standard model \cite{modelop}. The interaction of neutrinos with the Earth can be measured through the attenuation of the incident neutrino flux that crosses the Earth and is measured by the IceCube, as illustrated in Fig. \ref{fig:diagram}. 
The IceCube detector can measure High Energy Neutrino Sample, with energies above $60\;\mathrm{TeV}$. In this energy range the predominant interaction is deeply inelastic scattering \cite{formaggio}. For the neutrino to reach the detector, it is necessary that it does not interact via charged current when crossing the Earth. Neutral current interaction only decreases the energy of the beam. Although in this work we focus on the analysis of the interaction of muon neutrino, the results are also very similarly applicable for electron neutrino. In the energy limit of the neutrino much larger than the mass of the lepton produced, electron and  muon neutrinos have the same cross section with hadronic targets. However, for the analysis of the absorption of electron neutrino it is necessary to also include the effects of interaction with electrons (Glashow resonance) \cite{glashow,gandhi}. For recent work on neutrino absorption considering Glashow resonance, see Refs. \cite{IceCube2,victor,victor2}.

\section{Formalism}
\label{Sec:Form}

A correct description of the proton spin from quarks is of particular interest for neutrino physics. Neutrino and antineutrino are left-handed and right-handed chiral eigenstates, respectively. 
So, (anti)neutrino capturing the (right)left-hand component of the quark wavefunction, and any change between distribution of the quarks right and left-hand in the nucleons will impact specifically (anti)neutrino scattering and absorption cross section. 
At this point one must notice that besides the fact that weak neutrino-nucleon interactions are given in terms of chiral states. Thus, the information about how much of the nucleon spin is due to the quarks, {\it i. e.},  the value of $\Delta \Sigma(Q^{2})$, is not taken into account in the most common procedure of calculation of neutrino-nucleon cross-section at  the deep inelastic regime \footnote{At the high neutrino energies, where the description of the nucleon target in terms of form factors is no longer available.}. 
At sufficiently high energies, due to  {\it asymptotic freedom} \cite{Gross:1973id,Politzer:1973fx}, it is possible to describe the neutrino-nucleon interaction in terms of the neutrino scattering on free quarks that constitute the nucleon. The assumption of the  equal distribution of left and right quark spins leads to the average of the initial polarization state. As both the {\it charged current (CC)} and {\it neutral current (NC)} DIS  processes  are inclusive reactions,  they take into account all the possible final hadronic states, and a sum over all  the final state polarization possibilities  is also applied \cite{halzen}.  In \cite{pol}, the formalism to include spin effects at (anti)neutrino-nucleon interaction is presented.

In the  unpolarized DIS (CC), the neutrino $\nu_l$ (antineutrino $\bar{\nu}_l$) with energy $E_\nu$ interacts by exchanging a virtual boson $W^\pm$ of four - momentum $q$ ($q^2 = -Q^2$). The initial lepton becomes the associated charged lepton $l^{\pm} (=e, \mu, \tau)$ with energy $E'$ and the hadron goes to an unknown state of invariant mass $W$, characterized by $W^2>m_N^2$, with $m_N$ being the mass of the hadronic target. In terms of Bjorken's $x$, the inelasticity $y=(E_\nu-E')/E_\nu$ and the virtuality of the exchanged boson $Q^2$, the unpolarized double differential  cross section of neutrino DIS is given by \cite{reno}
\begin{eqnarray}
\begin{aligned}
\frac{\mathrm{d}\sigma^{\nu(\bar{\nu})}}{\mathrm{d}x\mathrm{d}y}=
\frac{G_F^2E_\nu m_N}{\pi}
\left( \frac{M_W^2}{Q^2+M_W^2} \right)^2
\left\{
\left(
y^2x+\frac{m_l^2y}{2E_\nu m_N}
\right)F_1(x,Q^2)+\right. 
\left(
1-y-\frac{m_l^2}{4E_\nu^2}-\frac{m_Nxy}{2E_\nu}
\right)F_2(x,Q^2)+ \\
 +(-)\left(
xy-\frac{xy^2}{2}-\frac{m_l^2y}{4E_\nu m_N}
\right)F_3(x,Q^2)
\left. +\frac{m_l^2 (m_l^2+Q^2)}{E_\nu^2 m_\nu^2 x}F_4(x,Q^2)-\frac{m_l^2}{E_\nu m_N}F_5(x,Q^2)
\right\} 
  \,\,\, ,
\label{Eq:Diff_cross_section}
\end{aligned}
\end{eqnarray}
where $G_F$ is the Fermi's constant, $M_W$ the $W^\pm$ boson mass, $m_l$ the mass of the lepton produced and $F_i$ are the spin independent structure functions. In the parton model, $F_2(x,Q^{2})$ is interpreted in terms of the sum of the helicity distributions of the quarks that have the flavor that can interact with the neutrino \cite{paschos}. $F_1(x,Q^{2})$ can be written in terms of $F_2(x,Q^{2})$ with the Callan-Gross relation and $F_3(x,Q^{2})$ is associated with quark - antiquark asymmetry. In this paper we use the CTEQ18 parameterization \cite{cteq18} for quark distributions, which uses the DGLAP evolution equations. In the high energy limit, which we are interested in, we also assume that the Albright - Jarlskog relations \cite{aj} hold. The standard variables of DIS are connected by $Q^2=2E_\nu m_Nxy$, in the rest frame of target. 
%Henceforth, the integrated neutrino-nucleon cross-section is given by
%\begin{eqnarray}
% \sigma^{\nu(\bar \nu)} = \int^{x_{0}}_{x_{f}} \int^{y_{0}}_{y_{f}}\frac{\mathrm{d}\sigma^{\nu(\bar{\nu})}}{\mathrm{d}x\mathrm{d}y}dxdy.
%\end{eqnarray}

%The inclusion of the dependence on $m_{l}$ also implies that the limits of integration are no longer $0\le x \le 1$ and $ 0 \le y \le 1$. Stead they are given in \cite{reno} as

%\begin{eqnarray}
%\begin{aligned}
% \frac{m^{2}_{l}}{2m_{N}(E_{\nu}-m_{l})}\le &x& \le 1 ~;~~~~~a-b\le y \le a+b ~; ~~~~~~~~%\mbox{where} 
%\end{aligned}
%\end{eqnarray}

%\noindent where

%\begin{eqnarray}
%\begin{aligned}
% a & = & \frac{1-m^{2}_{l}\left( \frac{1}{2m_{N}E_{\nu}x}  + \frac{1}{2E^{2}_{\nu}}    \right)}{2\left(1 + \frac{m_{N}x}{2E_{\nu}}  \right)}~;  \\
% b & = & \frac{\sqrt{\left( 1 - \frac{m^{2}_{l}}{2m_{N}E_{\nu}x}  \right) -\frac{m^{2}_{l}}{E^{2}_{\nu}}  } }{2\left(1 + \frac{m_{N}x}{2E_{\nu}}  \right)}.  \\
%\end{aligned}
%\end{eqnarray}

%Let us define the polarized cross-section difference
%\begin{eqnarray}
% \Delta \sigma \equiv \sigma(\lambda_{N} = -1 ) - \sigma(\lambda_{N} = +1 ).
%\end{eqnarray}
Moreover, when we consider the polarized hadronic target, the cross section of the Eq. \ref{Eq:Diff_cross_section} is modified by a factor dependent on the hadronic spin, given by \cite{pol,pol2}
\begin{eqnarray}
\begin{aligned}
\frac{\mathrm{d}\Delta\sigma^{\nu(\bar{\nu})}}{\mathrm{d}x\mathrm{d}y} = \frac{G_F^2ME_\nu}{\pi}\left( \frac{M_W^2}{Q^2+M_W^2}\right)^2 \lambda_N \left\{ (-) \left[ - yx(2-y)+\frac{2 x^3y^3m_l^2}{Q^2}\right] \right. g_1(x,Q^2)  
+ (-)\left( \frac{4x^3y^2m_l^2}{Q^2} \right)g_2(x,Q^2)+ \\
+\frac{2xym_l^2}{Q^2}\left( 1-y-\frac{x^2y^2m_l^2}{Q^2} \right)g_3(x,Q^2)
+\left[ -1+y-\frac{2x^2ym_l^2}{Q^2}\left( 1-\frac{3y}{2}-\frac{x^2y^2m_l^2}{Q^2} \right) \right]g_4(x,Q^2)+ \\
\left. +\left( -y^2x+\frac{2x^4y^3m_l^2}{Q^2} \right)g_5(x,Q^2) \right\}
\,\,\, ,
\label{correction}
\end{aligned} 
\end{eqnarray}
where $\lambda_N$ is the helicity of the hadronic target and $g_i(x,Q^{2})$ are the spin - dependent structure functions. Unlike the $F_i(x,Q^{2})$ functions, the $g_i(x,Q^{2})$ functions in the parton model are written with the difference of the helicity distributions of the each quarks, describing the net amount of quarks with spin in a given direction \cite{pol2}.  In the limit of energies of interest in this work, we assume the validity of the Dicus relation \cite{dicus}. Such a relation allows us to write $g_4$ in terms of $g_5$, similar to the Callan-Gross relation, $g_{4}(x,Q^{2}) = 2xg_{5}(x,Q^{2})$.   Both relationships emerged from the observation that, when we neglect the masses involved, helicity is conserved in the quark-gluon coupling. 
The polarized structure functions are described in \cite{pol}, where it is shown that the contributions from $g_{2}(x,Q^{2})$ and $g_{3}(x,Q^{2})$ to the cross-sections are suppressed by factors like $m^{2}_{l}/Q^{2}$, which could be measured in the neutrino factories. However, in this work we are  initially interested in $Q^{2}\ge \approx m^{2}_{W}$, which is the typical value for the neutrino-nucleon interactions at the IceCube neutrino observatory. Hence, we can disregard contributions from $g_{2}(x,Q^{2})$ and $g_{3}(x,Q^{2})$. Also, in the same limit, as the same structures appears in in both polarized and unpolarized cases, the polarized contribution to the  neutrino-nucleon cross-section can be obtained from Eq. \ref{Eq:Diff_cross_section} replacing $F_{1}(x,Q^{2}) \rightarrow - g_{5}(x,Q^{2})$,   $F_{2}(x,Q^{2}) \rightarrow - g_{4}(x,Q^{2})$, and $F_{3}(x,Q^{2}) \rightarrow 2 g_{1}(x,Q^{2})$. 
%The leading-order expression of $g_{1}(X,Q^{2})$ and $g_{5}(X,Q^{2})$ are then \cite{pol,pol2, Ridolfi:1999cf}:

%\begin{eqnarray}
%g^{W^{+}}_{1}(x,Q^{2}) &=& \Delta \bar u(x,Q^{2}) + \Delta d(x,Q^{2}) + \Delta \bar c(x,Q^{2}) + \Delta s(x,Q^{2}) \nonumber \\
%g^{W^{-}}_{1}(x,Q^{2}) &=& \Delta u(x,Q^{2}) + \Delta \bar d(x,Q^{2}) + \Delta  c(x,Q^{2}) + \Delta \bar s(x,Q^{2}) \nonumber \\
%g^{W^{+}}_{5}(x,Q^{2}) &=& \Delta \bar u(x,Q^{2}) - \Delta d(x,Q^{2}) + \Delta \bar c(x,Q^{2}) - \Delta s(x,Q^{2}) \nonumber \\
%g^{W^{-}}_{5}(x,Q^{2}) &=& -\Delta u(x,Q^{2}) + \Delta \bar d(x,Q^{2}) - \Delta  c(x,Q^{2}) + \Delta \bar s(x,Q^{2})
%\end{eqnarray}

\begin{figure}[t]
	\centering
			\includegraphics[width=0.6\textwidth]{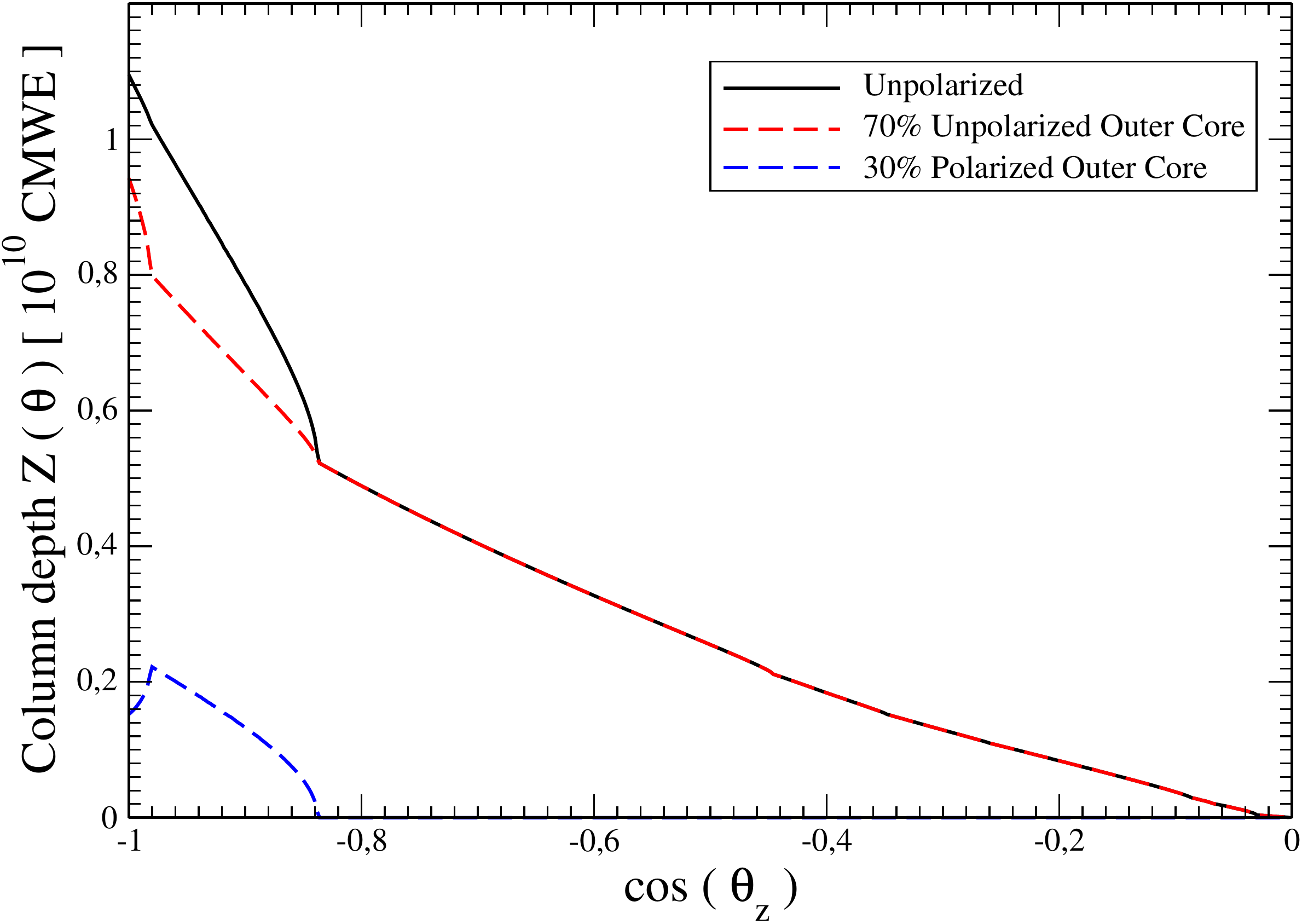}   
	\caption{The thickness of the Earth in centimeters of the water for nucleons unpolarized and with outer core partially polarized. We use the PREM model \cite{prem}.}
	\label{fig:column}
\end{figure}

An accurate description of the neutrino - nucleon cross section, as well as the distribution of matter in the interior of the Earth, is fundamental to estimate the absorption of neutrinos that cross the Earth, given that these are fundamental ingredients for the calculation. The probability of the neutrino crossing without being absorbed can be quantified with \cite{gandhi}
\begin{eqnarray}
P_{Shad}(E_\nu ,\theta_z) = 
\mathrm{exp}\left[
-N_A\sigma (E_\nu)\int^{r(\theta_z)}_{0}\rho_N(r)\mathrm{d}r
\right]
\,\,\, ,
\label{absorption}
\end{eqnarray}
where $N_A$ is the Avogadro numbers, $\rho(r)$ the Earth's density profile and $r(\theta_z) = -2R_{Earth}\mathrm{cos}\; \theta_z$ is the total distance travelled by neutrino. In this work we use the PREM model \cite{prem} for the description of the Earth's profile density. In Fig. \ref{fig:column} we show the thickness of matter traversed by the neutrino as a function of the zenith angle. We see two distinct cases: in the continuous black line we show the thickness of matter crossed without considering polarization. While the dashed curves show the thickness of matter traversed unpolarized (red) and polarized (below) considering a hypothetical case of 30\% polarization in the outer core (blue). The PREM model indicates that the outer core is located between 1221.5 km and 3480.0 km, which implies that the neutrino crosses this potentially polarized layer only if it hits with $\mathrm{cos}\;(\theta_z)$ less than $-0.84$.

\begin{figure}[b]
	\centering
		\begin{tabular}{ccc}
			\includegraphics[width=0.45\textwidth]{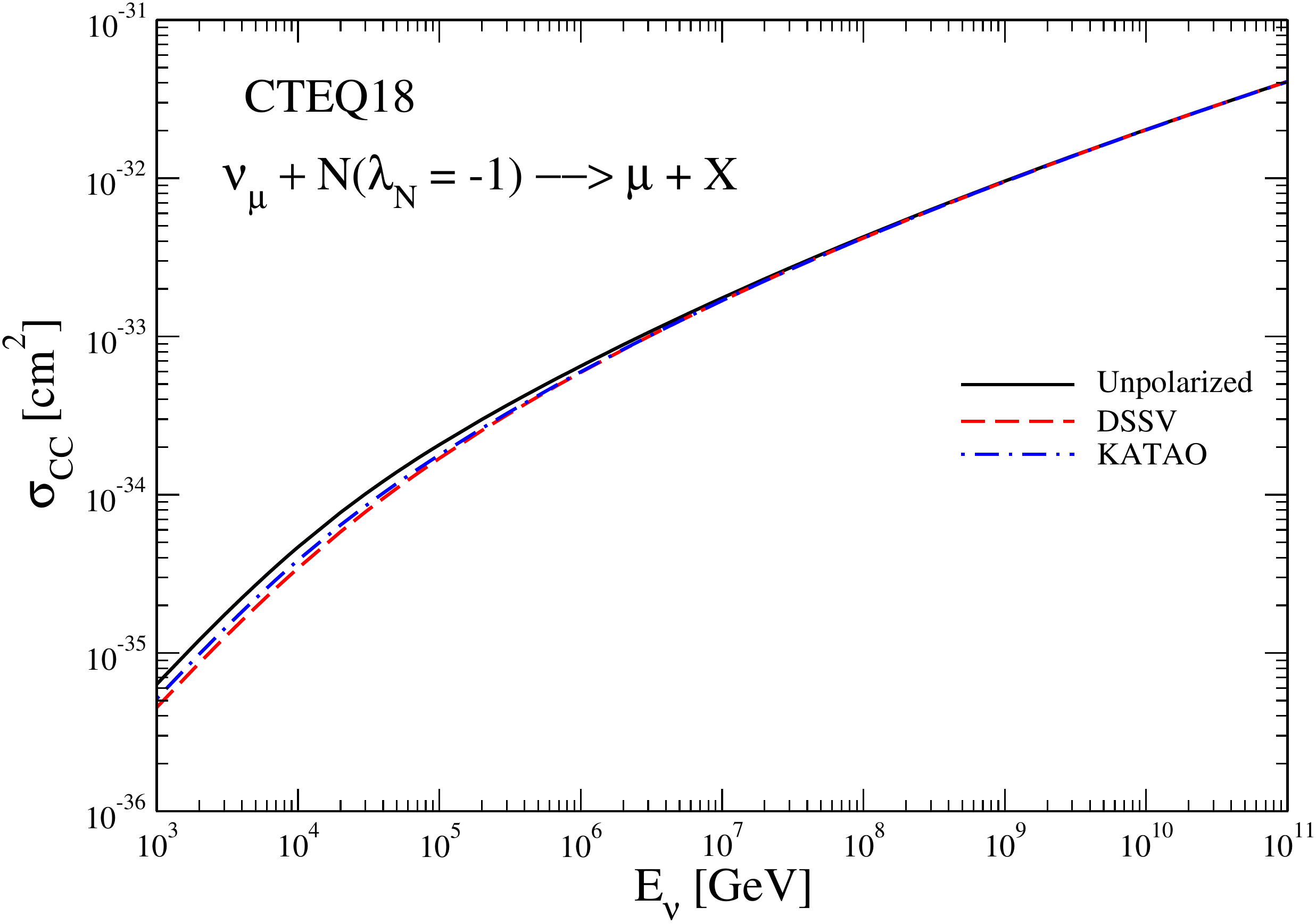} & \,\,\,\,\, &
			\includegraphics[width=0.45\textwidth]{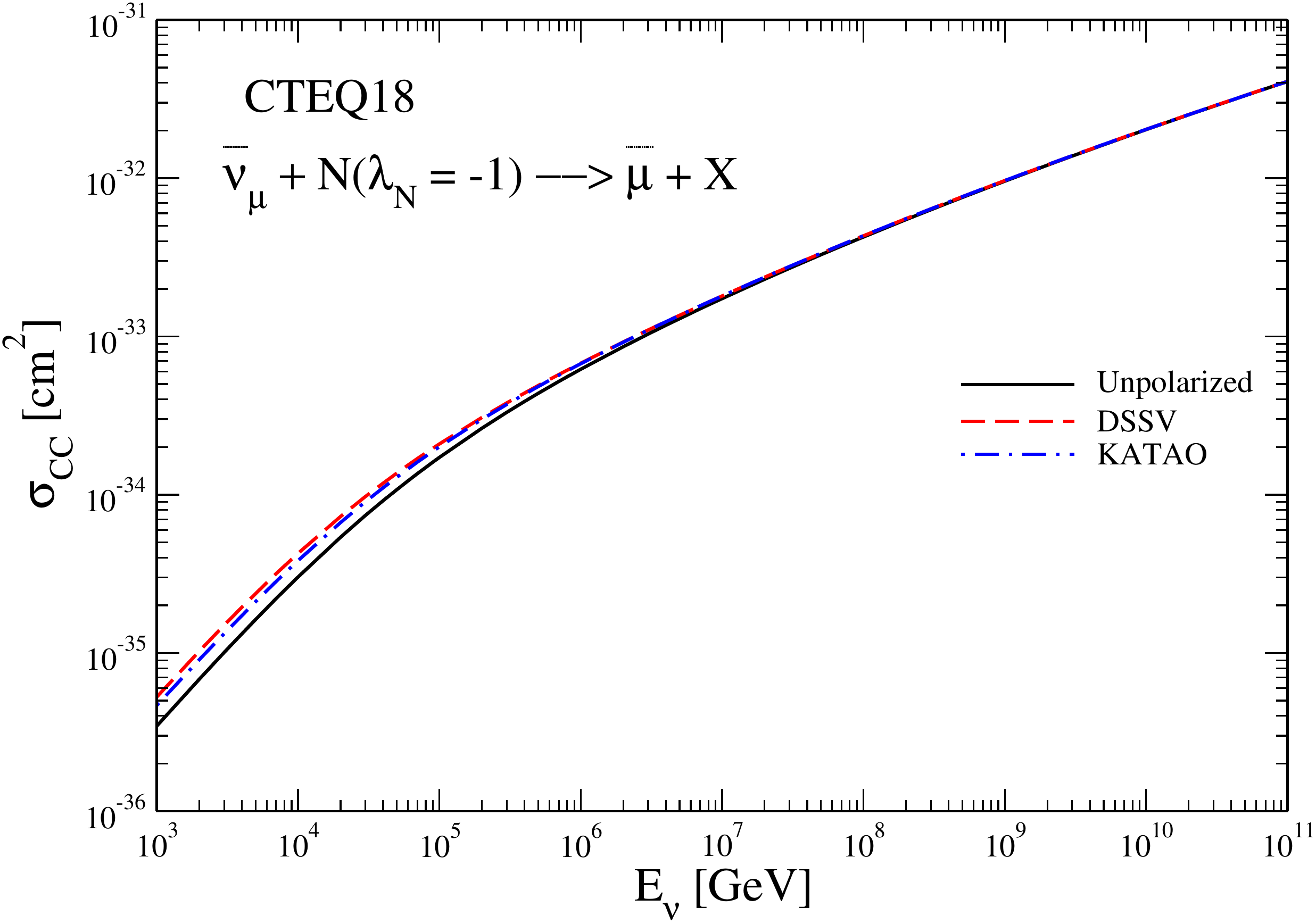} \\
			(a) & \, &  (b)   \\
			\includegraphics[width=0.45\textwidth]{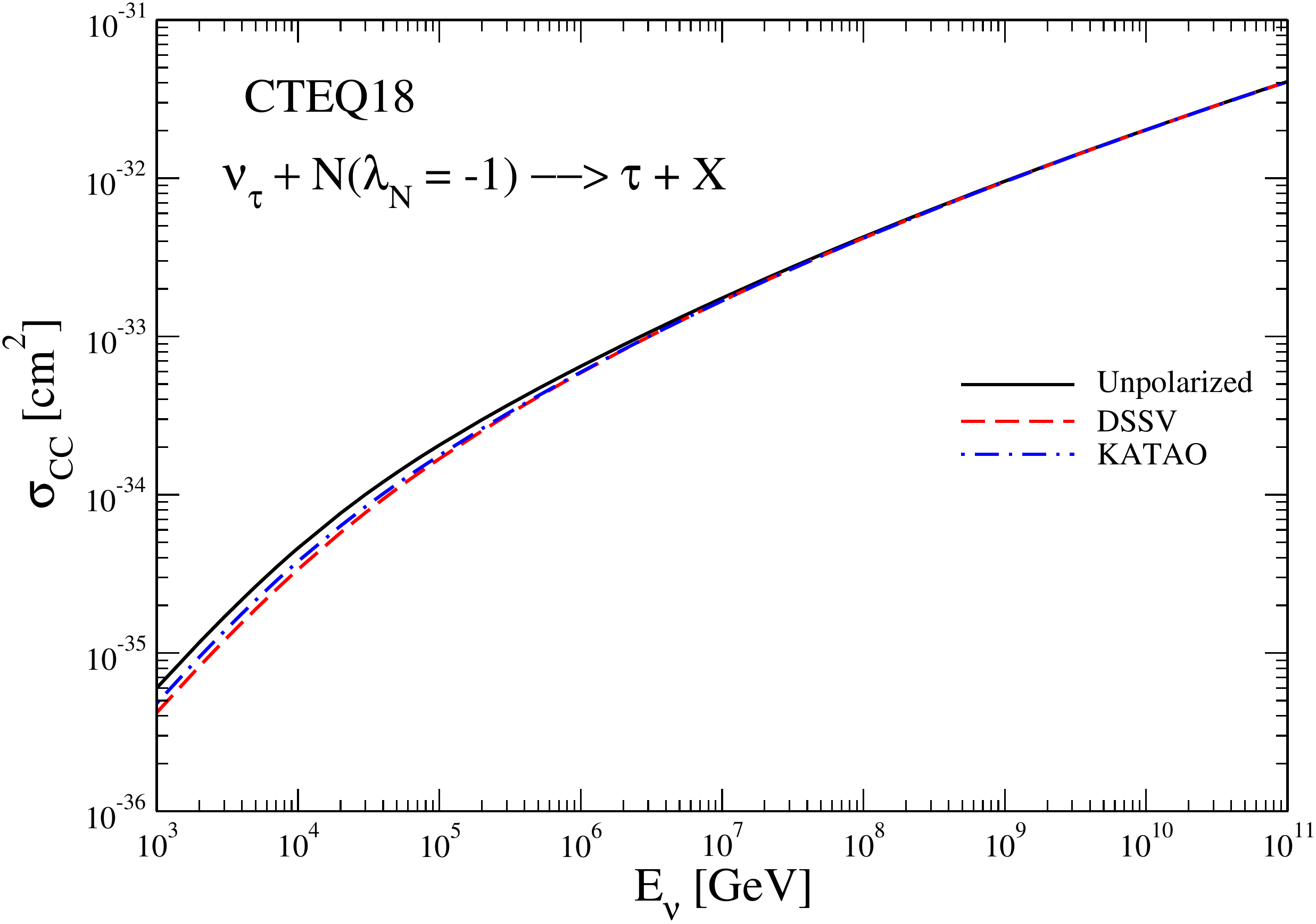} & \,\,\,\,\, &
			\includegraphics[width=0.45\textwidth]{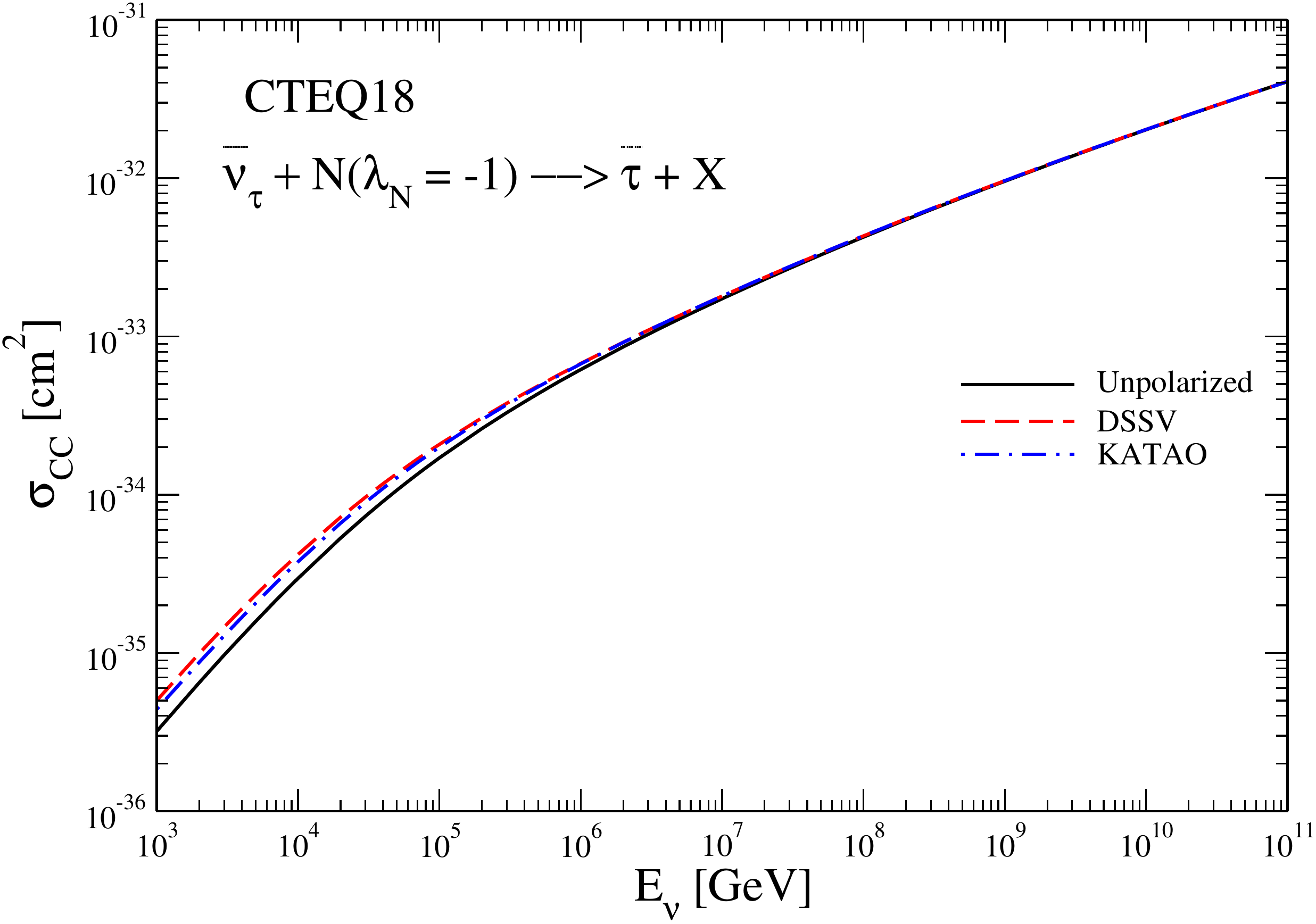} \\
			(c) & \, &  (d)   
		\end{tabular}
	\caption{Cross section for muon (a) neutrino and (b) antineutrino DIS with the isoscalar target. In (c) and (d) we present the same results for tau neutrino. We consider two distinct cases: unpolarized target and polarized with DSSV \cite{florian} and  KATAO \cite{KATAO} parametrizations of spin - dependent structure functions.	}
	\label{fig:cross_section}
\end{figure}

\section{Results}

Initially, we present in Fig. \ref{fig:cross_section} the cross section of (a) muon and (c) tau neutrino; and (b) muon and (d) tau antineutrino with a isoscalar target as a function of (anti)neutrino energy. We calculate the cross sections for unpolarized  and  polarized targets with helicity $\lambda_N = -1$. For the calculation of the polarized cross sections, we use two different parameterizations of the spin-dependent structure functions, DSSV \cite{florian} and KATAO \cite{KATAO}. Both parameterizations lead to similar results. The impact of target polarization on the cross section becomes less significant with increasing energy, practically disappearing for energies above $10^7\;\mathrm{GeV}$. For lower energies of incoming neutrino, $ 10^3 - 10^4\; \mathrm{GeV}$, the unpolarized and polarized cross sections differ by a multiplicative factors of $0.7 - 1.3$, depending on the parameterization of the spin - dependent structure functions and whether the beam is neutrino or antineutrino.

\begin{figure}
	\centering
		\begin{tabular}{ccc}
			\includegraphics[width=0.45\textwidth]{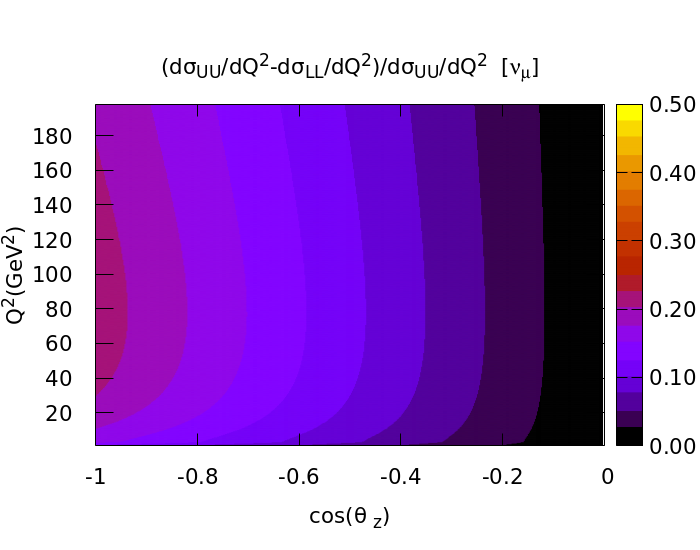} & \,\,\,\,\, &
			\includegraphics[width=0.45\textwidth]{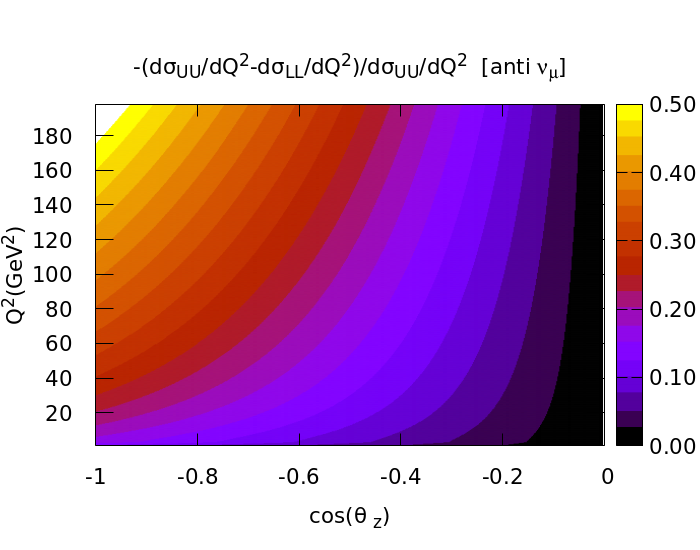} \\
			(a) & \, &  (b)   
		\end{tabular}
	\caption{Difference between unpolarized and polarized differential cross sections normalized by unpolarized differential cross sections for (a) neutrino and (b) antineutrino incident. We calculate using the KATAO \cite{KATAO} and CTEQ18 \cite{cteq18} parameterization, for neutrino and antineutrino incident with energies of $10^3\; \mathrm{GeV}$.  }
	\label{fig:ratio}
\end{figure}

\begin{figure}%[t]
	\centering
		\begin{tabular}{ccc}
			\includegraphics[width=0.45\textwidth]{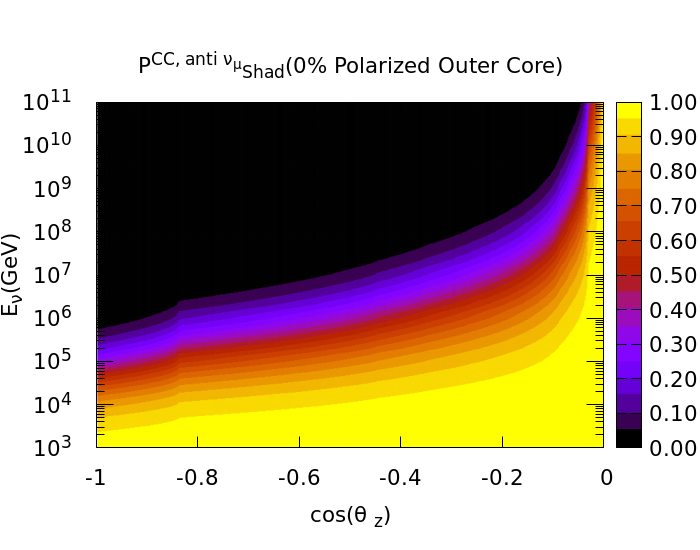} &
			\includegraphics[width=0.45\textwidth]{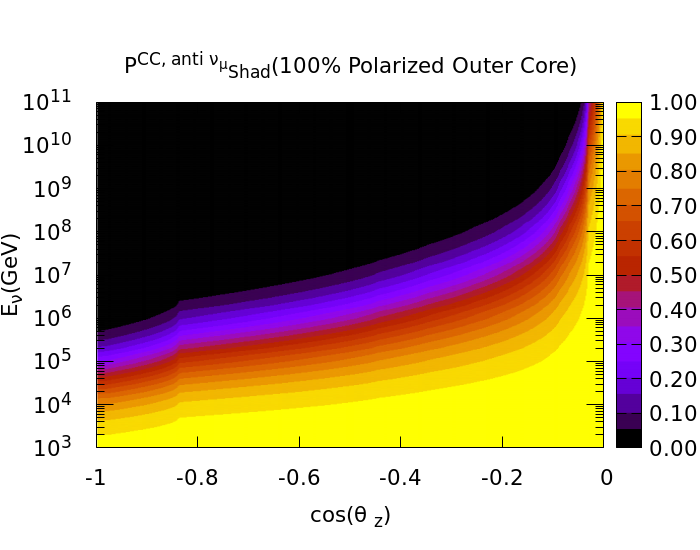} \\
			(a) & (b) \\
			\includegraphics[width=0.45\textwidth]{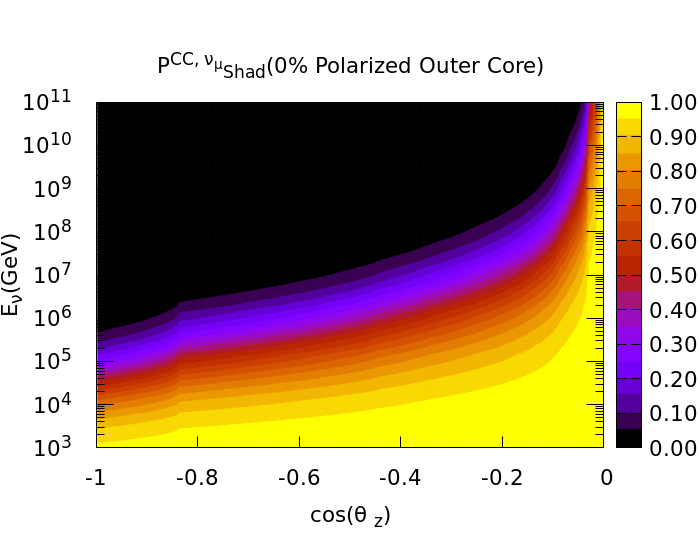} & 
			\includegraphics[width=0.45\textwidth]{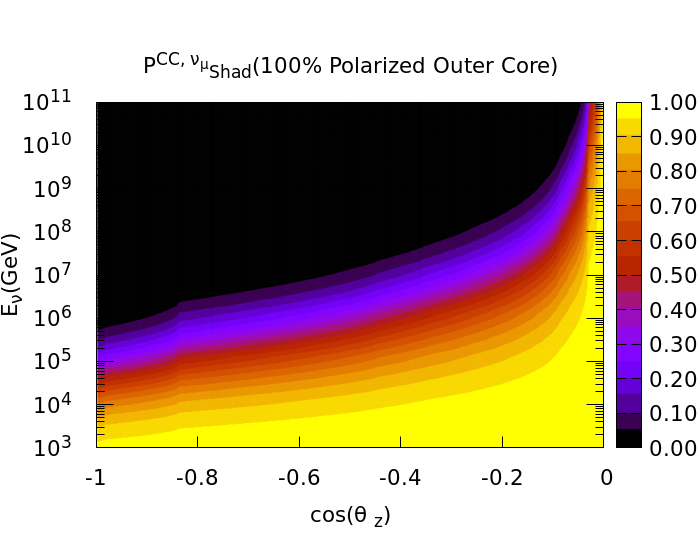} \\
			(c) & (d)
		\end{tabular}
	\caption{Probability of the  antineutrino, (a) and (b), and  neutrino, (c) and (d), crossing the Earth without interacting via charged current as a function of energy and $\mathrm{cos}\; (\theta_z)$ considering the  unpolarized Earth, (a) and (c), and  with 100\% polarization in the outer core, (b) and (d). }
	\label{fig:P_abs}
\end{figure}

To better trace the origin of the polarization effect on the nutrino-nucleon interaction (CC), we present Fig. \ref{fig:ratio}. In it we quantify the difference between the unpolarized (UU) and polarized (LL) differential cross section, normalized with the unpolarized differential cross section. $\theta_z$ is the angle between the direction of arrival of the neutrino (antineutrino) with the spin of the isoscalar target, considering again $\lambda_N = -1$. We clearly see that the difference between the differential cross sections is maximized when the spins are parallel or antiparallel to the direction of propagation of the incident neutrino. In $Q^2$, this normalized difference is maximized in different regions for neutrino and antineutrino. While for the neutrino we have a maximum of  $\approx 80\; \mathrm{GeV}^2$, for the antineutrino the maximum goes beyond $200\; \mathrm{GeV}^2$, the region with the smallest contribution to the total cross section.

In Fig. \ref{fig:P_abs} we present our result for the probability of crossing without being absorbed, $P_{Shad}$, of muonic neutrinos by the Earth as a function of the energy and the cosine of the zenith angle of incidence of the neutrino. In the upper panel, we present $P_{Shad}$ for antineutrinos and in the lower panel for neutrinos. In the Figs. \ref{fig:P_abs}a and \ref{fig:P_abs}c  we do not consider any polarization on earth. In the Figs. \ref{fig:P_abs}b and \ref{fig:P_abs}d,  $P_{Shad}$ is calculated considering 100\% polarization in the outer core of the Earth. The calculation of the cross sections disregarding polarization is performed with the structure functions constructed with the quark distributions parameterized by CTEQ18 \cite{cteq18}. For the cross section with polarization correction of the hadronic target (Figs. \ref{fig:P_abs}b and \ref{fig:P_abs}d), we used the KATAO  parameterization \cite{KATAO} spin - dependent structure functions, besides, of course, CTEQ18 for the spin - independent structure functions. The choice of KATAO parameterization for this result is practically indifferent to the choice of DSSV, because, as previously discussed and illustrated in Fig. \ref{fig:cross_section}, both lead to very similar cross sections.

\begin{figure}[t]
	\centering
		\begin{tabular}{ccc}
			\includegraphics[width=0.45\textwidth]{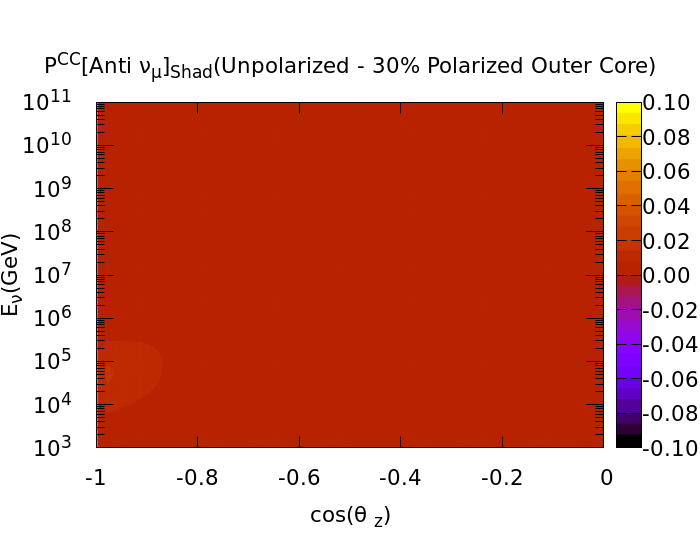} &
			\includegraphics[width=0.45\textwidth]{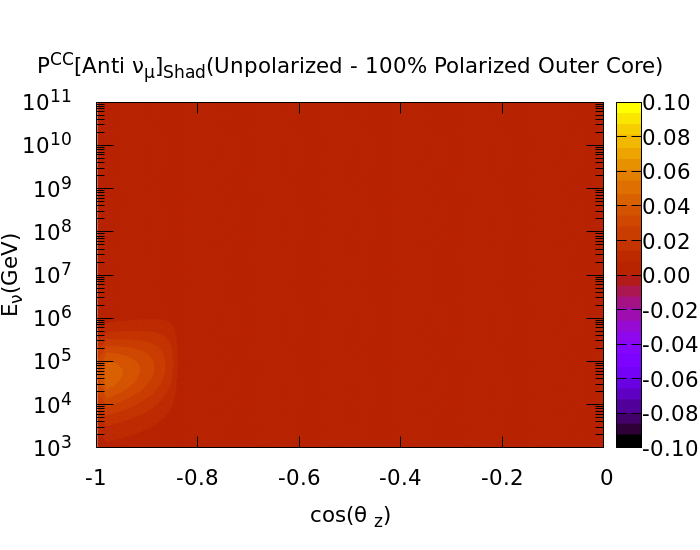} \\
			(a) & (b) \\
			\includegraphics[width=0.45\textwidth]{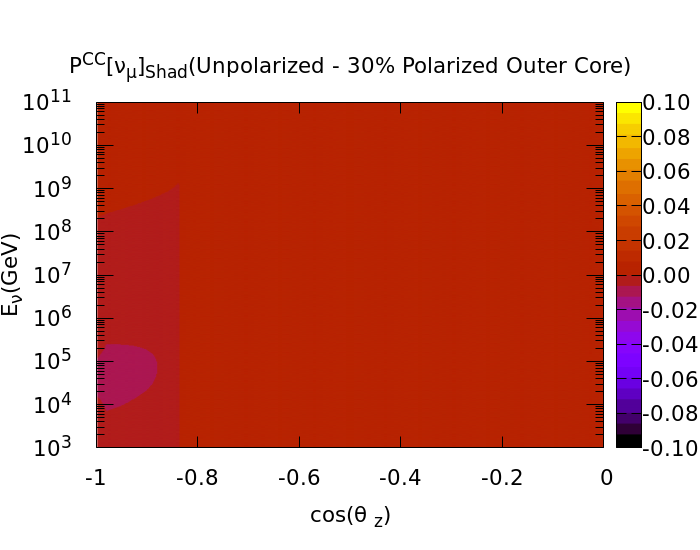} & 
			\includegraphics[width=0.45\textwidth]{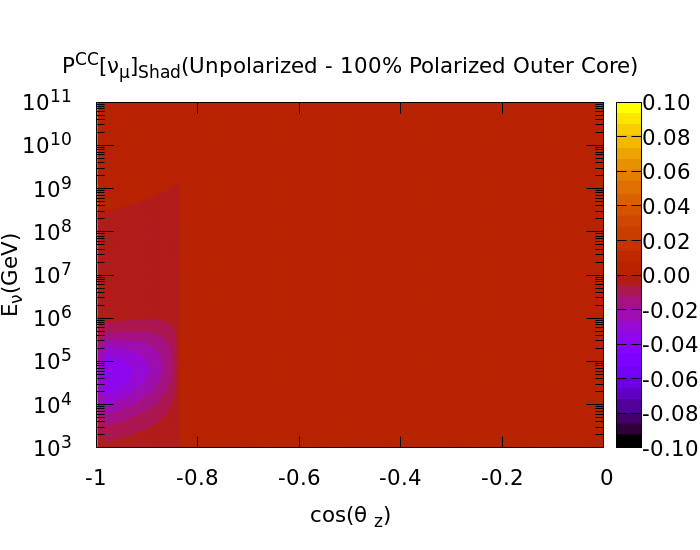} \\
			(c) & (d)
		\end{tabular}
	\caption{Difference between probability of the  antineutrino, (a) and (b), and  neutrino, (c) and (d), crossing the Earth without interacting via charged current as a function of energy and $\mathrm{cos}\; (\theta_z)$ considering the  unpolarized Earth, (a) and (c), and  with 100\% polarization in the outer core, (b) and (d). }
	\label{fig:diff_P_abs}
\end{figure}

In Fig. \ref{fig:diff_P_abs} we present the difference between the absorption probabilities of unpolarized and polarized Earth. In Figs. \ref{fig:diff_P_abs}a and \ref{fig:diff_P_abs}c (\ref{fig:diff_P_abs}b and \ref{fig:diff_P_abs}d) are the results for the absorption of antineutrino (neutrino) considering two cases: 30\% and 100\% of polarization in the outer core, respectively. The effect is restricted to the region where the neutrino impinges with $\mathrm{cos}\; (\theta_z) < -0.84$, given that this is the necessary condition for it to cross the outer core. We can see that neutrino and antineutrino have opposite effects: while $P_{Shad}$ of neutrinos decreases with the effect of polarization, $P_{Shad}$ of antineutrinos increases with said effect. This fact hinders the experimental validation of the model, since the IceCube does not distinguish the charge of the produced lepton. Even considering 100\% polarization the effect is small, however in the IceCube observation region ($10^3 - 10^6\;\mathrm{GeV}$).

To better quantify the difference described above and presented in Fig. \ref{fig:diff_P_abs}, we calculate the percentage difference between unpolarized and polarized absorptions ($(P_{Shad}^{UNPOL.}-P_{Shad}^{POL.})/P_{Shad}^{UNPOL.}$). We estimate that in the IceCube observation region where the absorption is significant, with 30\% polarization the absorption changes between 0\% and $\pm$5\%. For 100\% polarization this percentage rises to about $\pm$18\%. It is still possible to observe that the mentioned effect is maximized with $\mathrm{cos}\; (\theta_z) \rightarrow -0.98$, when the neutrino crosses the largest possible amount of the outer core. Despite being significant percentages in the change of absorption, it is very difficult to observe it. Because we use high percentages of polarization to get it, and the effects on neutrinos and antineutrinos are of similar but opposite magnitudes. To observe this effect, a future detector capable of distinguishing between neutrinos and antineutrinos would be needed.

\section{Summary}

In this study we have investigated the impact of the polarization of the Earth's outer core on the absorption of neutrinos in the IceCube observation region. 
Our main motivation was to verify if a possible estimate of the polarization of the Earth's outer core can be made in the future by attenuating the flux of neutrinos that cross the Earth. 
Our results showed that the effects of polarization on absorption, although in the IceCube region, are small even considering 100\% polarization. 
Added to this, while the absorption of neutrinos decreases, that of antineutrinos increases with the polarization of the hadronic targets, making it a difficult task to estimate the nuclear polarization by the attenuation of the neutrino flux. 
Given the magnitude of the impact of polarization on neutrino absorption, a more detailed analysis of the angular and energy distributions of the expected number of events in the IceCube detector becomes unfeasible. 
The future EIC may change the current view we have of the contribution of sea quarks to the proton spin, and motivate more detailed analyzes of neutrino absorption considering polarization on Earth. 
Furthermore, with the neutrino detectors of the future, such as IceCube-Gen2 \cite{IceCubeGen2} and GRAND \cite{grand}, the prospect of measuring these smaller magnitude effects may become feasible. 
This work is far from a comprehensive estimate of possible spin QCD effects.
Even if earth polarization is exactly zero, it does not mean spin dependent effects vanish:   It is known that correlations of spin in nuclei are strong, i.e. each nucleon's polarization depends on the others.   ``One level down", parton level correlations are both strong and relatively unknown (for instance, is the up quark more likely to be aligned or anti-aligned with the proton's spin?).   
Accounting for these effects would require convoluting into the earth's profile both the nuclear spin wavefunction of iron and medium-modified quark TMDs, and is currently beyond the scope of this work. 
 We show the impact of polarization only on muon neutrinos absorption. However, the effect on electronic neutrinos is essentially the same, as they have the same cross section in the high energy limit. For tauonic neutrinos there is a greater difference in cross section at lower energies, due to the mass of the tau produced. However, a more detailed analysis of tau neutrinos absorption for future detectors presupposes the study of flux regeneration by tau decay, which is outside the scope of this work.

\begin{acknowledgments}
This work was  partially financed by the Brazilian funding
agencies CNPq and CAPES. G.T.~acknowledges support from Bolsa de produtividade CNPQ
306152/2020-7, Bolsa de pesquisa FAPESP 2021/01700-2, Partecipation
to Tematico FAPESP, 2017/05685-2 and the
grant BPN/ULM/2021/1/00039 from the Polish National Agency for Academic Exchange.
\end{acknowledgments}

\hspace{1.0cm}


\begin{thebibliography}{99}

\bibitem{EMC_1988}
J.~Ashman {\it et al.} [EMC Collaboration],
  %``A measurement of the spin asymmetry and determination of the structure function $g_1$ in deep inelastic muon-proton scattering,''
  Phys.\ Lett.\ B {\bf 206}, 364 (1988);
  
  
  
 
  \bibitem{EMC_1989}J.~Ashman {\it et al.}  [EMC Collaboration],
  %``An investigation of the spin structure of the proton in deep inelastic scattering of polarized muons on polarized protons,''
  Nucl.\ Phys.\ B {\bf 328}, 1 (1989);


 %\cite{Collins:1989gx}
\bibitem{Collins:1989gx}
J.~C.~Collins, D.~E.~Soper and G.~F.~Sterman,
%``Factorization of Hard Processes in QCD,''
Adv. Ser. Direct. High Energy Phys. \textbf{5}, 1-91 (1989);
%doi:10.1142/9789814503266\_0001
%[arXiv:hep-ph/0409313 [hep-ph]].
%1511 citations counted in INSPIRE as of 23 Feb 2023 
  

  

%\cite{Workman:2022ynf}
\bibitem{Workman:2022ynf}
R.~L.~Workman \textit{et al.} [Particle Data Group],
%``Review of Particle Physics,''
PTEP \textbf{2022}, 083C01 (2022);
%doi:10.1093/ptep/ptac097
%3 citations counted in INSPIRE as of 12 Jul 2022   


  \bibitem{pol}
  S.~Forte, M.~L.~Mangano and G.~Ridolfi,
  %``Polarized parton distribution from charged-current deep-inelastic scattering and future neutrino factories,''
Nucl. \ Phys.\ B {\bf 602}, 585–621 (2001);




   \bibitem{pol2}
M.~Anselmino, P.~Gambino and J.~Kalinowski,
  %``Polarized deep inelastic scattering at high energies and parity violating structure functions,''
Zeit. \ Phys.\ C {\bf 64}, 267 (1994);



  
%\cite{Ridolfi:1999cf}
%\bibitem{Ridolfi:1999cf}
%G.~Ridolfi,
%``Polarized parton distributions in perturbative QCD,''
%Nucl. Phys. A \textbf{666}, 278-281 (2000).
%doi:10.1016/S0375-9474(00)00036-1
%[arXiv:hep-ph/9910485 [hep-ph]].
%4 citations counted in INSPIRE as of 25 Feb 2023
  
  


\bibitem{COMPASS}
C.~Adoiph {\it et al.} [COMPASS Collaboration],
  %``Final COMPASS results on the deuteron spin-dependent structure function and the Bjorken sum rule,''
  Phys.\ Lett.\ B {\bf 769}, 34-41 (2017);

\bibitem{Spin_quark_sea}
C.~A.~Aidala, S.~D.~Bass, D.~Hasch and G.~K.~Mallot,
  %``The Spin Structure of the Nucleon,''
  Rev.\ Mod.\ Phys. {\bf 85}, 655 (2013);
  
  
 

\bibitem{COMPASS3}
V. Y. Alexakhin {\it et al.} [COMPASS Collaboration],
  %``The deuteron spin-dependent structure function g1d and its first moment,''
  Phys.\ Lett.\ B {\bf 647}, 8 (2007);

  
\bibitem{compass2}
C.~Adolph {\it et al.} [COMPASS Collaboration],
  %``The spin structure function $g_1$ of the proton and a test of the Bjorken sum rule,''
  Phys.\ Lett. \ B {\bf 753}, 18 (2016);  
  
\bibitem{clas}
R.~G.~Fersch {\it et al.} [CLAS Collaboration],
  %``Determination of the proton spin structure functions for $0.05<Q^@<5$ GeV$^2$ using CLAS,''
  Phys.\ Rev. \ C {\bf 96}, 065208 (2017);  


\bibitem{hermes2}
A.~Airapetian {\it et al.} [HERMES Collaboration],
  %``Precise determination of the spin structure function $g_1$ of the proton, deuteron and neutron,''
  Phys.\ Rev. \ D {\bf 75}, 012007 (2007);  

\bibitem{SMC}
B. Adeva {\it et al.} [Spin Muon Collaboration],
  %``Measurement of the spin-dependent structure function $g_1$ (x)$$ of the deuteron,''
  Phys.\ Lett. \ D {\bf 302}, 533 (1993);

\bibitem{e143}
K.~Abe {\it et al.} [E143 Collaboration],
  %``Measurements of the proton and deuteron spin structure functions $g_1$ and $g_2$,''
  Phys.\ Rev. \ D {\bf 58}, 112003 (1998);  

\bibitem{e142}
P. L. Anthony {\it et al.} [E142 Collaboration],
  %``Deep inelastic scattering of polarized electrons by polarized 3 He and the study of the neutron spin structure,''
  Phys.\ Rev. \ D {\bf 54}, 6620 (1996); 

\bibitem{e154}
K. Abe {\it et al.} [E142 Collaboration],
  %``Precision determination of the neutron spin structure function g1n,''
  Phys.\ Rev. \ Lett. {\bf 79}, 26 (1997);


  
\bibitem{jlab}
N.~Sato, W.~Melnitchouk, S.~E.~Kuhn, J.~J.~Ethier and A.~Accardi, [Jefferson Lab Angular Momentum Collaboration],
  %``Iterative Monte Carlo analysis of spin-dependent parton distributions,''
  Phys.\ Rev. \ D {\bf 93}, 074005 (2016);  

\bibitem{KATAO}
A.~N.~Khorramian, S.~Atashbar~Tehrani, S.~Taheri~Monfared, F.~Arbabifar and F.~I.~Olness,
  %``Polarized deeply inelastic scattering (DIS) structure functions for nucleons and nuclei,''
Phys.\ Rev. \ D {\bf 83}, 054017 (2011);


\bibitem{KATAO2}
H.~Khanpour, S.~Taheri~Monfared and S.~Atashbar~Tehrani,
  %``Polarized deeply inelastic scattering (DIS) structure functions for nucleons and nuclei,''
Phys.\ Rev. \ D {\bf 95}, 074006 (2017);


\bibitem{KATAO3}
H.~Khanpour, S.~Taheri~Monfared and S.~Atashbar~Tehrani,
  %``Nucleon spin structure functions, considering target mass correction and higher twist effects at the NNLO accuracy and their transverse momentum dependence,''
Phys.\ Rev. \ D {\bf 105}, 074023 (2022);


 \bibitem{florian}
D.~de~Florian, R.~Sassot, M.~Stratmann and W.~Vogelsang,
  %``Extraction of spin-dependent parton densities and their uncertainties,''
 Phys.\ Rev. \ D {\bf 80}, 034030 (2009); 


\bibitem{florian2}
I.~Borsa, D.~de~Florian and I.~Pedron.
  %``The full set of Polarized Deep Inelastic Scattering Structure Functions at NNLO accuracy,''
  	arXiv:2210.12014 [hep-ph]; 


\bibitem{Cocuzza:2022ovz}
C.~Cocuzza \textit{et al.} [JAM],
%``Sea Asymmetry from Polarized $W$ Boson Production,''
SciPost Phys. Proc. \textbf{8}, 025 (2022);
%doi:10.21468/SciPostPhysProc.8.025 
  
  %\cite{Ethier:2017zbq}
\bibitem{Ethier:2017zbq}
J.~J.~Ethier, N.~Sato and W.~Melnitchouk,
%``First simultaneous extraction of spin-dependent parton distributions and fragmentation functions from a global QCD analysis,''
Phys. Rev. Lett. \textbf{119} no.13, 132001 (2017);
%doi:10.1103/PhysRevLett.119.132001
%[arXiv:1705.05889 [hep-ph]].
%171 citations counted in INSPIRE as of 15 Feb 2023

\bibitem{lattice}
Y.~Yang {\it et al.}
  %``Glue Spin and Helicity in the Proton from Lattice QCD,''
  Phys.\ Rev.\ Lett. {\bf 118}, 102001 (2017);
  
\bibitem{exp_gluon}
D.~de~Florian, R.~Sassot, M.~Stratmann and W.~Vogelsang,
  %``Evidence for Polarization of Gluons in the Proton,''
  Phys.\ Rev.\ Lett. {\bf 113}, 012001 (2014);



%\cite{Scopetta:1999ue}
\bibitem{Scopetta:1999ue}
S.~Scopetta and V.~Vento,
%``Orbital angular momentum parton distributions in quark models,''
[arXiv:hep-ph/9907441 [hep-ph]];
%0 citations counted in INSPIRE as of 17 Feb 2023

\bibitem{Ratcliffe}
 P. G. Ratcliffe  %Orbital angular momentum and the parton model. 
Phys. Lett. B, v. 192, n. 1-2, p. 180-184, (1987);



%\cite{Gribov:1972ri}
\bibitem{Gribov:1972ri}
V.~N.~Gribov and L.~N.~Lipatov,
%``Deep inelastic e p scattering in perturbation theory,''
Sov. J. Nucl. Phys. \textbf{15}, 438-450
IPTI-381-71 (1972);
%4984 citations counted in INSPIRE as of 17 Feb 2023

%\cite{Altarelli:1977zs}
\bibitem{Altarelli:1977zs}
G.~Altarelli and G.~Parisi,
%``Asymptotic Freedom in Parton Language,''
Nucl. Phys. B \textbf{126}, 298-318 (1977).
%doi:10.1016/0550-3213(77)90384-4
%7882 citations counted in INSPIRE as of 17 Feb 2023
%\cite{Dokshitzer:1977sg}

\bibitem{Dokshitzer:1977sg}
Y.~L.~Dokshitzer,
%``Calculation of the Structure Functions for Deep Inelastic Scattering and e+ e- Annihilation by Perturbation Theory in Quantum Chromodynamics.,''
Sov. Phys. JETP \textbf{46}, 641-653 (1977);
%4574 citations counted in INSPIRE as of 17 Feb 2023






 
  

\bibitem{introduction_gpd}
M.~Diehl,
  %``Introduction to GPDs and TMDs,''
  Eur.\ Phys.\ J. \ A {\bf 52}, 149 (2016);  
  
\bibitem{introduction_gpd2}
C.~Riedl,
  %``Probing nucleon spin structure in deep-inelastic scattering, proton-proton collisions and Drell-Yan processes,''
  Acta\ Phys.\ Pol. \ B {\bf 53}, 5-A2 (2022);  


  

\bibitem{hermes}
A.~Airapetian {\it et al.} [HERMES Collaboration],
  %``Measurement of the proton spin structure function $g_1^p$ with a pure hydrogen target,''
  Phys.\ Lett. \ B {\bf 442}, 484 (1998);  
  


  
\bibitem{eic}
A.~Accardi  {\it et al.}
  %``Electron-Ion Collider: The next QCD frontier,''
 Eur. \ Phys.\ J. \ A {\bf 52}, 268 (2016); 
 
 

  	


 
\bibitem{geodynamics}
N.~Schaeffer, D.~Jault, H.~C.~Nataf and A.~Fournier,
  %``Turbulent geodynamo simulations: a leap towards Earth's core,''
  Geophys.\ J.\ Int. {\bf 211}, 1–29 (2017); 
  
  
  \bibitem{measure}
M.~G.~Aartsen  {\it et al.} [IceCube],
  %``Measurement of the multi-TeV neutrino interaction cross-section with IceCube using Earth absorption,''
 Nature {\bf 551}, 596-600 (2017); 
 
   \bibitem{modelop}
A.~Cooper-Sarkar, P.~Mertsch and S.~Sarkar,
  %``The high energy neutrino cross-section in the Standard Model and its uncertainty,''
 JHEP {\bf 1108}, 042 (2011); 
 
 \bibitem{formaggio}
J.~A.~Formaggio and G.~P.~Zeller,
  %``From eV to EeV: Neutrino cross sections across energy scales,''
  Rev.\ Mod.\ Phys. {\bf 84}, 1307 (2012);


  \bibitem{glashow}
S.~L.~Glashow,
  %``Resonant Scattering of Antineutrinos,''
   Phys.\ Rev. {\bf 118}, 316 (1960);
   
   \bibitem{gandhi}
  R.~Gandhi, C.~Quigg, M.~H.~Reno and I.~Sarcevic,
  %``Ultrahigh-Energy Neutrino Interactions,''
  Astropart.\ Phys.\ D {\bf 5}, 81 (1996);
  
  
  \bibitem{IceCube2}
M.~G.~Aartsen  {\it et al.} [IceCube],
  %``Detection of a particle shower at the Glashow resonance with IceCube,''
 Nature {\bf 591}, 220 (2021), [Erratum: Nature {\bf 592}, E11 (2021)]; 
 
 
  \bibitem{victor}
V.~P.~Gonçalves, D.~R.~Gratieri and A.~S.~C.~Quadros,
  %``Estimating the impact of the QCD dynamics on the determination of the high energy astrophysical neutrino flux,''
   Eur.\ Phys.\ J. \ C {\bf 81}, 496 (2021);
   
    \bibitem{victor2}
V.~P.~Gonçalves, D.~R.~Gratieri and A.~S.~C.~Quadros,
  %``Implications of the QCD dynamics and a Super-Glashow astrophysical neutrino flux on the description of ultrahigh energy neutrino data,''
   Eur.\ Phys.\ J. \ C {\bf 82}, 1011 (2022);



  
%\cite{Gross:1973id}
\bibitem{Gross:1973id}
D.~J.~Gross and F.~Wilczek,
%``Ultraviolet Behavior of Nonabelian Gauge Theories,''
Phys. Rev. Lett. \textbf{30}, 1343-1346 (1973);
%doi:10.1103/PhysRevLett.30.1343
%5978 citations counted in INSPIRE as of 24 Feb 2023  
  

%\cite{Blumlein:1998nv}
%\bibitem{Blumlein:1998nv}
%J.~Blumlein and A.~Tkabladze,
%``Target mass corrections for polarized structure functions and new sum rules,''
%Nucl. Phys. B \textbf{553} (1999), 427-464
%doi:10.1016/S0550-3213(99)00289-8
%[arXiv:hep-ph/9812478 [hep-ph]].
%153 citations counted in INSPIRE as of 25 Feb 2023  
  
  
  
%\cite{Politzer:1973fx}
\bibitem{Politzer:1973fx}
H.~D.~Politzer,
%``Reliable Perturbative Results for Strong Interactions?,''
Phys. Rev. Lett. \textbf{30}, 1346-1349 (1973);
%doi:10.1103/PhysRevLett.30.1346
%5750 citations counted in INSPIRE as of 24 Feb 2023

  

\bibitem{halzen}
F.~Halzen and A.~D.~Martin,
  %``Quark & Leptons: An Introductory Course In Modern Particle Physics,''
  John Wiley $\&$ Sons, New York, (2008);  

  
  

 
 \bibitem{reno}
S.~Kretzer and M.~H.~Reno,
  %``Tau neutrino deep inelastic charged current interactions,''
 Phys.\ Rev.\ D {\bf 66}, 113007 (2002);

 \bibitem{paschos}
E.~A.~Paschos and J.~Y.~Yu,
  %``Neutrino interactions in oscillation experiments,''
 Phys.\ Rev.\ D {\bf 65}, 033002 (2002);


  \bibitem{cteq18}
  Tie-Jiun Hou {\it et al.}
  %``New CTEQ global analysis of quantum chromodynamics with high-precision data from the LHC,''
  Phys.\ Rev.\ D {\bf 103},  014013 (2021); 
  
  
  
   \bibitem{aj}
C.~H.~Albright and C.~Jarlskog,
  %``Neutrino production of M$^+$ and E$^+$ heavy leptons (I),''
  Nucl. \ Phys.\ B {\bf 84}, 467 (1975);
  
  

  

  \bibitem{dicus}
  D.~A.~Dicus,
  %``Relations Among the Structure Functions of Deep-Inelastic Neutrino-Nucleon Scattering,''
  Phys.\ Rev.\ D {\bf 5}, 1367 (1972);
  

  
\bibitem{prem}
  A~.M.~Dziewonski and D~.L.~Anderson,
  %``Preliminary reference Earth model,''
  Phys.\ Earth Plan. \ Int. {\bf 25}, 297 (1981);

  
  \bibitem{IceCubeGen2}
M.~G.~Aartsen  {\it et al.} [IceCube-Gen2],
  %``IceCube-Gen2: the window to the extreme Universe,''
 J.\ Phys.\ G {\bf 48},     060501 (2021); 

  
  \bibitem{grand}
J.~Álvarez-Muñiz  {\it et al.} [GRAND Collaboration],
  %``The Giant Radio Array for Neutrino Detection (GRAND): Science and Design,''
 Sci.\ China Phys.\ Mech.\ Astron. {\bf 63}, 219501 (2020); 




\end{thebibliography}
\end{document}